\documentclass[aps,prl,twocolumn,showpacs,groupedaddress,superscriptaddress,amssymb]{revtex4}

\usepackage{graphicx}

\newcommand{\eqref}[1]{Eq.~(\ref{#1})}
\newcommand{\figref}[1]{Fig.~\ref{fig:#1}}
\newcommand{\deffig}[4]{
\begin{figure}[b]
  \begin{center}
  \includegraphics[width=#3 \textwidth]{#2}
  \end{center}
  \caption{ \label{fig:#1} #4}
\end{figure}
}

\newcommand{\Ham}{{\cal H}}
\newcommand{\Eq}[1]{(\ref{#1})}

\begin{document}

\title{Transition matrix Monte Carlo method for quantum systems}

\author{Chiaki Yamaguchi}
\affiliation{
Department of Computer and Mathematical Sciences,
 Graduate School of Information Sciences, Tohoku University,
 Aramaki-Aza-Aoba 04, Aoba-ku, Sendai 980-8579, Japan
}

\author{Naoki Kawashima}
\affiliation{
  Department of Physics, Tokyo Metropolitan University,
  Tokyo 192-0397, Japan
}

\author{Yutaka Okabe}
\affiliation{
  Department of Physics, Tokyo Metropolitan University,
  Tokyo 192-0397, Japan
}

\date{\today}

\begin{abstract}
We propose an efficient method for Monte Carlo simulation
of quantum lattice models.
Unlike most other quantum Monte Carlo methods,
a single run of the proposed method yields 
the free energy and the entropy with high precision 
for the whole range of temperature.
The method is based on several recent findings
in Monte Carlo techniques, such as the loop algorithm 
and the transition matrix Monte Carlo method.
In particular, we derive an exact relation
between the DOS and the expectation value of the 
transition probability for quantum systems, 
which turns out to be useful in reducing the 
statistical errors in various estimates.
\end{abstract}


\maketitle



\section{Introduction}

The Monte Carlo method for classical and quantum lattice models 
has been improved dramatically since the proposal of 
the Metropolis method\cite{Metropolis}.
The introduction of the extended ensemble,
being one of the ideas that enhanced the Monte Carlo method,
is characterized by a random walker traveling in the energy space.
The most well-known methods among the ones based on the extended ensemble
is the multicanonical method \cite{BergN,Lee}.
In this method, one can obtain the density of states (DOS) 
from the histogram of visiting frequency at each value of the energy.
Since the random walk in the extended ensemble methods is typically
biased in a complicated way, the scaling property of the 
the traveled distance of the walker deviates from
that of the free random walk, i.e., $R \sim \sqrt{t}$.
This means that the random-walk nature of the method does not 
necessarily guarantee a quick equilibration.
However, in many important application, it turns out that
the random walker can still travel the whole range of the energy
within a reasonable computational time, i.e., a computational
time bounded by some polynomial in the system size.
Therefore, the method is quite useful for models that
has a first-order phase transition and/or ones that have 
local minima in the free energy.
In addition, the direct calculation of the DOS makes
it possible to calculate the absolute value of the free energy 
of a given system with high precision.
Another advantage of this type of approach is
that thermal averages of various quantities can be obtained for the 
whole range of the temperature from only a single run of simulation.
While the early versions of extended ensemble methods still
suffered from a slow diffusion,
even this problem was removed recently by
the acceleration method proposed by Wang and Landau (WL) \cite{WangL},
in which the random walker is forced to move away whenever 
it lingers about a place for a long time.

For many important models, classical or quantum,
it was found that the loop/cluster algorithm
\cite{SwendsenW,Evertz} is so efficient that the simulation can be
done with no (or negligible) difficulty of slowing down at any temperature.
However, since the algorithm is originally designed for a simulation at a fixed 
temperature, one cannot obtain from a single run 
the whole temperature dependence of
various quantities nor can one estimate the free energy.
Therefore, it is natural to wonder if the loop/cluster algorithm
can be used together with the extended ensemble methods.
An apparent obstacle is that the energy in the 
quantum case is not defined so that constructing a
histogram on it is straightforward and meaningful.
Janke and Kappler proposed\cite{JankeK} that
in the cluster algorithm for classical spin systems
one can use the number of graph elements (such as bonds
in the Swendsen-Wang algorithm)
instead of the energy for constructing the histogram,
showing a way for combining the extended ensemble method 
and the cluster algorithm for classical spin systems.
Furthermore, two of the present authors 
demonstrated\cite{YamaguchiK} the enhancement of the efficiency 
achieved with the use of the WL method,
and also suggested that the resulting framework can be 
generalized to quantum systems.
Later, the application to the quantum systems was 
done with a different formulation
by Troyer, Wessel and Alet \cite{TroyerWA},
who also showed the efficiency of the resulting algorithm.

In the present paper, we propose that a further improvement
can be achieved for quantum models
by using the broad-histogram (BH) relation for estimating the DOS.
The idea has been partially described in our previous paper
\cite{YamaguchiKO} for the classical models.
For the Ising models,
it was pointed out \cite{Oliveira1} that a ratio of the
densities of states at two neighboring values of the energy
can be expressed in terms of the microcanonical averages of
the numbers of distinct potential moves.
In the application to the Ising model,
the number of potential moves is nothing but the number of
spins such that flipping one of them causes a change in the
energy by a given amount.
Since this is a macroscopic quantity, whereas the
number of the visits at each value of the energy is not,
we can estimate the DOS more precisely by using the relation 
than we do so directly from the histogram.
In the case that we discuss here, 
our DOS is not simply the total number of states.
Therefore we have to use the generalized form of
 the transition matrix Monte Carlo method \cite{WangTS}.
In what follows, we describe the method specialized for
quantum systems.

\section{The outline of the algorithm}
The simulation consists of two stages.
In the first stage we perform short runs
using the WL method for obtaining a working estimate of the DOS.
As explained below, each run in the first stage is characterized
by a parameter called the modification factor, which controls 
the adoptive change of the fictitious weight of the random walk.
A stronger control is imposed on earlier runs whereas 
a weaker control for the latter runs.
Because of this adoptively changing weight,
the microcanonical ensemble obtained in each run of the first stage 
deviates from the correct one.
This leads to some systematic error.
However, we need unbiased estimates of microcanonical averages\cite{Note1} 
because we use them for obtaining the DOS from the BH relation.
Therefore, in the second stage we perform a single long run
with the weight fixed to the one obtained in the first stage.
This is in contrast to the ordinary WL method where the outcome of the
first stage is taken as the final estimate of the DOS.
The organization of the whole simulation is schematically
depicted in \figref{Simulation}.
We will show this two-stage procedure gives more precise results
for quantum systems than the one-stage procedure of the WL method
described in \cite{TroyerWA}, within the same
amount of the total CPU time.
\deffig{Simulation}{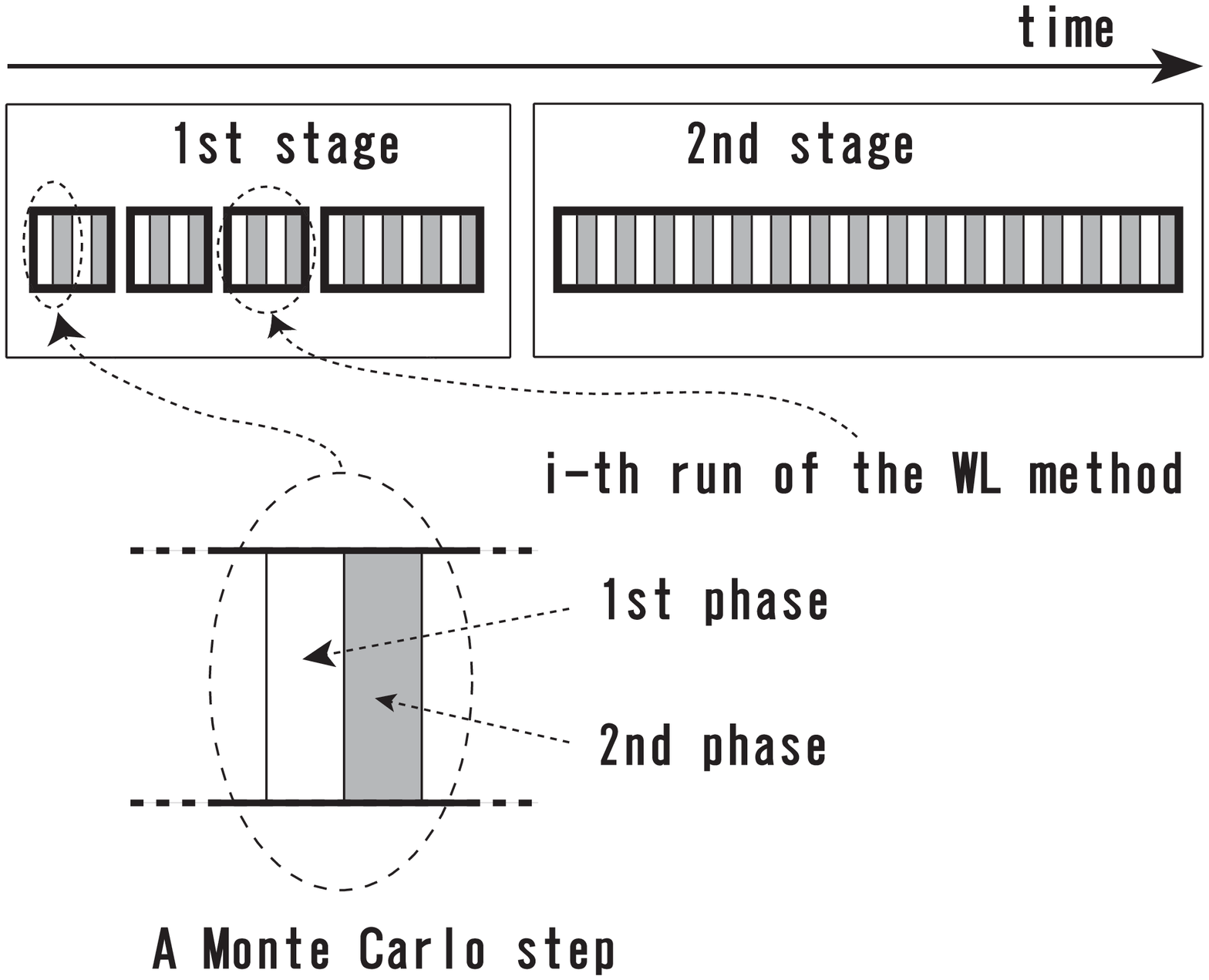}{0.45}
{The organization of a set of simulation.}

A little more specifically, 
the whole simulation can be described as follows.
We construct a histogram of the expansion order or the number
of active vertices, which we denote as $n$, 
instead of the histogram of the energy.
Accordingly, 
the fictitious weight $f(n)$ is introduced as a function of 
$n$, rather than the energy.
The Monte Carlo dynamics is defined 
so that the detailed balance holds for this weight.
We start the first run of the first stage by setting the weight $f(n)$ 
to be constant, i.e., $f(n) = 1$ for all $n$.
At the beginning of each run, 
the counter (or the histogram), $H(n)$, of visits to the value $n$
is set to be 0 while $f(n)$ is not reset except at the beginning of the
first run.
For the $i$-th run we use the modification 
factor $\lambda_i \equiv e^{-2^{-i}}$.
Every time the random walker visit a state with $n(S) = n$,
$f(n)$ is replaced by $\lambda_i f(n)$
in order to penalize the random walker staying at this value of $n$.
At the same time, $H(n)$ is increased by one.
The $i$-th run is terminated when the histogram becomes flat.
To be specific, the flatness condition is
$$
  H(n) \ge 0.8 \times \sum_{n'} H (n') / \sum_{n'} 1
$$
for all $n$.
Since the first stage is only for obtaining a working estimate of 
the DOS, we stop when $- \ln \lambda_i < 10^{-5}$
 whereas in the original WL method
 $- \ln \lambda_i < 10^{-8}$ was used instead.
At the end of the first stage, $(f(n))^{-1}$ is approximately 
proportional to the generalized DOS discussed below.
In the second stage, we do the same as the first stage but 
with a longer run and with $f(n)$ fixed to be the result 
of the first stage.
In this way, we can compute non-biased microcanonical averages of
various quantities in the second stage.

In both the first and the second stages, 
the update of the configuration is done using the loop update
if the update does not involve a change in $n$.
Other similar algorithms such as the directed loop algorithm
\cite{SS} can be used as well, and the generalization is
straight-forward.
The choice, however, is only of the secondary importance 
for the applications to the models without the magnetic field 
as considered in the present paper.
For the loop algorithm, there are two formulations;
one is based on the continuous imaginary time and the other is
the discrete imaginary time.
Although the difference in the efficiencies of two formulations is minor,
in the present paper we have used the discrete time formulation
following Troyer, Wessel and Alet\cite{TroyerWA}.
In what follows, we reformulate their algorithm,
use it for deriving the BH relation, and describe how the relation 
is used in the second stage of the simulation.

The systematic error due to the discretization in the discrete
time formulation can be removed by starting from the high-temperature 
series expansion \cite{Sandvik} rather than the path-integral representation.
Therefore, we start with expressing the partition function $Z$ as
a high-temperature series expansion truncated at the $L$-th order,
\begin{eqnarray}
Z_0
& = & 
{\rm Tr} e^{-\beta\Ham}
\approx \sum_{n=0}^{L} \frac{\beta^n}{n!} {\rm Tr} \left(-\Ham\right)^n 
\nonumber \\
& = & 
\sum_{n=0}^{L} \beta^n\frac{L!}{(L-n)!} 
\sum_{\{\gamma_l\}\atop(\sum \gamma_l = n)} 
{\rm Tr} \prod_l (-\Ham)^{\gamma_l} 
\nonumber \\
& = & \sum_{n=0}^{L} \beta^n \, \omega(n) \, , 
\label{partition} 
\end{eqnarray}
where $\beta$ is inverse temperature $1 / k_B T$,
$k_B$ is Boltzmann's constant, and $\gamma_l$ is a variable
that takes on $0$ or $1$.
The high-temperature expansion coefficient, $\omega(n)$, 
is here regarded as the generalized DOS for the series order $n$.
When the Hamiltonian is expressed as a sum of 
$M$ local interaction terms as $\Ham = \sum_{b=1}^{M}\Ham_b$,
the generalized DOS is given by
\begin{equation}
  \omega(n) 
  =
    \frac{(L-n)!}{L!} 
    \sum_{
    \{ \gamma_{l,b} \} 
    \atop \left(
    {\sum_{l,b} \gamma_{l,b} = n
    \atop 
    \sum_b\gamma_{l,b} \le 1
    } \right) }
    \sum_{\{\psi_l\}} \prod_{l,b} \langle \psi_{l+1}|(-\Ham_b)^{\gamma_{l,b}}|\psi_l \rangle.
\end{equation}
We can simplify the notation by defining a {\it vertex}, $v$, as 
a pair of indices $l$ and $b$, i.e., $v=(l,b)$.
Several other symbols are also introduced for simplification of the
notation:
$\Gamma \equiv \{ v\,|\,\gamma_v = 1 \}$,
$\Psi \equiv (\psi_1, \psi_2, \cdots, \psi_L)$,
$\Psi_v \equiv (\psi_{l+1}, \psi_l)$,
$u(\Psi_v) \equiv \langle \psi_{l+1} | (-H_b) | \psi_l \rangle$,
$S \equiv (\Psi,\Gamma)$, and 
$n(S) \equiv |\Gamma|$.
As a result, we can rewrite the generalized DOS as
\begin{equation}
  \omega(n) = \sum_{{ S\atop (n(S)=n)} } U(S) \label{eq:Omega}
\end{equation}
where
\begin{equation}
  U(S) \equiv \frac{(L-n(S))!}{L!} \prod_{v \in \Gamma} u(\Psi_v)
  \prod_{v \notin \Gamma} \Delta(\Psi_v). \label{eq:U}
\end{equation}
Here, $\Delta(\Psi_v)\equiv \delta_{\psi_{l+1},\psi_{l}}$.
By comparing these expressions to those in \cite{YamaguchiK}
it is natural to identify $n$ as the number of graph elements.
Here we call a vertex $v$ for which $\gamma_v = 1$ an {\it active} vertex.

We then replace the factor $\beta^n$ in \Eq{partition} by 
an adjustable function $f(n)$, which makes the simulated partition function to be
$$
  Z = \sum_{n=0}^L f(n) \omega(n) = \sum_S W(S)
$$
with the weight of the state $S$ being
\begin{equation}
  W(S) = f(n(S)) U(S). \label{eq:W}
\end{equation}
In the first stage,
the function $f(n)$ is adaptively modified so that 
the probability of having $n(S)$ to be independent of $n$.
In other words, in the first stage
we try to make $f(n)\omega(n)$ constant by adjusting $f(n)$.
This goal is achieved by the use of the WL method.
As a result, we obtain $\omega(n)$ as the inverse of $f(n)$.

\section{The first phase and the BH relation}
The algorithm used for updating $S$ can be viewed as the dual Monte Carlo
\cite{KandelDomany,KawashimaG1995a}.
Namely, by regarding the variable $\gamma_v$ as a graph element,
analogous to the bond in the Swendsen-Wang algorithm,
one can establish a correspondence between the present algorithm and
the Swendsen-Wang algorithm.
One step in a dual Monte Carlo simulation consists of 
two phases; the ``graph'' $\Gamma$ is updated in the first phase
with ``spin configuration'' $\Psi$ being fixed
and the $\Psi$ is updated in the second phase with fixed $\Gamma$.
While the first phase changes $n$, the latter does not.
In what follows, we describe updating of $\Gamma$
and show how to estimate $\omega(n)$ based on estimates 
of macroscopic quantities.
The second phase, i.e., how the update $\Psi$ with fixed $n$,
is described in the next section.

In order to relate $\omega(n)$ to
the microcanonical averages of some macroscopic quantities,
we consider an arbitrary function $T(S'|S)$ that satisfies
\begin{equation}
  T(S'|S) W(S) = T(S|S') W(S'). \label{eq:DetailedBalance}
\end{equation}
While $T(S'|S)$ could be the transition probability from $S$ to $S'$,
in which case the above equation is nothing but the detailed balance
condition,
it does not have to be so for the derivation of the BH relation.
We sum up each side of the above equation
over such $S$ and $S'$ that $n(S) = n$ and $n(S') = n'$,
yielding
\begin{equation}
  \langle T(n'|S) \rangle_n f(n) \omega(n)
  =
  \langle T(n|S') \rangle_{n'} f(n') \omega(n')
  \label{eq:GeneralBHRelation}
\end{equation}
where
$$
  T(n'|S) \equiv \sum_{S'\atop (n(S') = n')} T(S',S),
$$
and $\langle \cdots \rangle_n$ is the microcanonical average, i.e.,
$$
  \langle Q(S) \rangle_n
  \equiv
  \left. \sum_{S\atop (n(S) = n)} Q(S) \right/ \sum_{S\atop (n(S) = n)} 1.
$$
For the derivation of the BH relation, we only have to consider the
case where $S'$ and $S$ have the same spin configuration $\Psi$ in common.
Then, the procedure for generating
$S'=(\Psi,\Gamma')$ from $S=(\Psi,\Gamma)$ is as follows:
\begin{description}
\item[1.]
    Generate an integer $l$ ($1\le l\le L$) uniform-randomly.
\item[2.]
    If $\gamma_l = 0$, 
    generate an integer $b$ ($1\le b\le M$) uniform-randomly, 
    and change $\Gamma$ into $\Gamma' \equiv \Gamma\cup \{v\}$
    (i.e., activate $v$)
    with the probability $p_v^+(S)$
    where $v\equiv (l,b)$.
\item[2'.]
    If $\gamma_l = 1$,
    change $\Gamma$ into $\Gamma' \equiv \Gamma \backslash \{v\}$ 
    (i.e., inactivate $v$)
    with the probability $p_v^-(S)$
    where $v$ is the vertex $(l,b)$ such that $\gamma_{l,b}=1$. 
\item[3.]
    Repeat 1 and 2 (or 2') $LM$ times.
\end{description}
The overall transition probability corresponding to one repetition
of 1 and 2 (or 2') can be written in the form of the transition matrix as
\begin{eqnarray*}
  T(S'|S) & = & 
  \frac1L \sum_l \Big(
  \delta_{\gamma_l,0} 
  \sum_b \frac1M p_{lb}^+(S) \Delta(\Gamma',\Gamma \cup \{(lb)\}) \\
  & &
  +\delta_{\gamma_l,1} 
  \sum_b \delta_{\gamma_{l,b},1} p_{lb}^-(S) 
  \Delta(\Gamma',\Gamma \backslash \{(lb)\})
  \Big),
\end{eqnarray*}
where some Kronecker's delta's have been
written as $\Delta(A,B)$ instead of $\delta_{A,B}$ for clarity.
In order for this transition probability to satisfy
\Eq{eq:DetailedBalance} with \Eq{eq:W} and \Eq{eq:U},
$p_v^{\pm}$ must satisfy
$$
  \frac1M (L-n) \Delta(\Psi_v) f(n) p_v^+(S) = u(\Psi_v) f(n+1) p_v^-(S')
$$
where $n\equiv n(S)$ and 
$v$ is the vertex at which $S$ and $S'$ differ from each other.
The simplest choice is
\begin{equation}
  p_v^+(S) \equiv \frac{M u(\Psi_v)}{L-n}\frac{f(n+1)}{f(n)}
  \quad\mbox{and}\quad p_v^-(S) \equiv \Delta(\Psi_v).
  \label{eq:ActivationProbability}
\end{equation}
Note here that we do not have to make $p_v^{\pm}$ smaller than 1
if we only need to make $T(S'|S)$ satisfy \Eq{eq:DetailedBalance}
and if $T$ does not have to be a real transition matrix.
Since this is the case for obtaining the BH relation, we shall
forget about the condition $p_v^{\pm} \le 1$ for a time being.
With \Eq{eq:ActivationProbability}, 
$T(n+1|S)$ and $T(n-1|S)$ become
$$
  T(n+1|S) = \frac1L \sum_l \delta_{\gamma_l,0} \sum_b \frac{u(\Psi_v)}{L-n}
  \frac{f(n+1)}{f(n)}
$$
and 
$$
  T(n-1|S) = \frac1L \sum_l \delta_{\gamma_l,1} \sum_b \delta_{\gamma_{l,b},1} \Delta(\Psi_v)
  = \frac1L (n-n_{\rm kink}),
$$
where $n_{\rm kink}$ is the number of kinks, i.e., the vertices at which
$\Delta(\Psi_v) = 0$.
Thus we obtain from \Eq{eq:GeneralBHRelation} for $n'=n+1$,
\begin{equation}
  \frac{\langle \sum_v \delta_{\gamma_l,0} u(\Psi_v) \rangle_n}{L-n}
  \, \omega(n) 
  =
  (n+1-\langle n_{\rm kink} \rangle_{n+1}) \, \omega(n+1) .
  \label{eq:BHRelation}
\end{equation}
We call this relation the BH relation for quantum systems.
When multiplied $\omega(n+1)/\omega(n)$ for $n=0,1,2,\cdots$, this yields
\begin{equation}
  \omega(n) = \frac{(L-n)!}{L!} \, 2^N \, \prod_{m=1}^{n} 
  \frac{\langle \sum_v \delta_{\gamma_l,0} u(\Psi_v) \rangle_{m-1}}
  {m-\langle n_{\rm kink} \rangle_{m}} \, ,
  \label{eq:TheEstimator}
\end{equation}
where $\omega(0)=2^N$ has been used.
Note that the validity of \Eq{eq:BHRelation} and \Eq{eq:TheEstimator}
does not depend on what transition probability
we use in the simulation although we have used a particular form of it
in deriving the relation \Eq{eq:BHRelation}.
Therefore, we can use any algorithm to estimate the  DOS
as long as it produces the correct microcanonical ensemble.

In fact, since we have not guaranteed that $p_v^+ \le 1$ above in
\Eq{eq:ActivationProbability}, $T(S'|S)$ mentioned above may not
serve as the transition probability.
To obtain the real transition probability, we simply replace 
\Eq{eq:ActivationProbability} by
\begin{eqnarray*}
  & & p_v^+(S) \equiv \min(1,R_{n(S)})  \quad\mbox{and}\quad \\
  & & p_v^-(S) \equiv \min(1,R_{n(S)-1}^{-1}) \Delta(\Psi_v),
\end{eqnarray*}
with
$$
  R_n \equiv \frac{M u(\Psi_v)}{L-n}\frac{f(n+1)}{f(n)}.
$$
This is the transition probability that is actually used in 
the simulation presented below.

\section{The second phase}
We now turn to the second phase in which
the state $S = (\Psi,\Gamma)$ is updated to $S' = (\Psi',\Gamma)$
with $\Gamma$ fixed.
This can be done by either of the loop algorithm or the
worm-type algorithm.
We here adopt the loop type update \cite{Sandvik99}.
The weight with which $\Psi$ must be sampled is $W(\Psi,\Gamma)$.
Following the general framework of the loop algorithm
we decompose the weight at every active vertex $v$ in $\Gamma$ as
\begin{equation}
  u(\Psi_v) = \sum_{G_v} a(G_v) \Delta(\Psi_v, G_v)
  \label{eq:Decomposition}
\end{equation}
where $G_v$ is a graph and $\Delta(\Psi_v,G_v)$ is the
function that takes on $0$ or $1$ depending on the 
matching between $\Psi_v$ and $G_v$.
The readers are referred to Ref.\cite{KawashimaG1995b} for
the details of the graphs and the $\Delta$-functions.
Here we only show an example for the compactness of the description.
In the case of the $S=1/2$ antiferromagnetic Heisenberg model,
the decomposition consists of a single term
since the local Hamiltonian can be expressed as a single delta function.
$$
  u(\Psi_v) = \frac{J}{2} \Delta(\Psi_v,G_v^{{\rm H}})
$$
where $G_v^{{\rm H}}$ is a {\it horizontal graph} that consists of
two horizontal lines connecting $(i,l)$ to $(j,l)$ and $(i,l+1)$ to $(j,l+1)$. 
The function $\Delta(\Psi_v,G_v^{{\rm H}})$ is defined
for $v=((ij),l)$ as
$$
  \Delta(\Psi_v,G_v^{{\rm H}})
  \equiv
  \left\{
  \begin{array}{ll}
  1 & (\psi_l(i)+\psi_l(j) = \psi_{l+1}(i)+\psi_{l+1}(j) = 0) \\
  0 & (\mbox{Otherwise})
  \end{array}
  \right.
  .
$$
Here we have assumed that the state $|\psi_l\rangle$ is the simultaneous
eigenstate of all the $z$-components of spins.
Namely, $\psi_l(i)$ $(=\pm 1/2)$ denotes the eigenvalue of $S_i^z$ of 
the eigenstate $|\psi_l\rangle$.
Therefore, the update is done by placing horizontal graphs on
all the vertices in $\Gamma$, and
connect the end points of horizontal lines by vertical lines
to form loops, then finally flip every loop with probability 1/2.
The generalization to the case where the expansion of the Hamiltonian
\Eq{eq:Decomposition} consists of multiple terms is straight-forward.
In such a case, we simply choose a graph from the ones for which
$\Delta(\Psi_v,G_v) = 1$ with the probability proportional to $a(G_v)$.


\section{Results}

In order to see if the present procedure yields better statistics
than previous methods,
we calculate the spin-1/2 antiferromagnetic 
Heisenberg chain of ten sites.
We set the initial value of the modification factor
to $e^{-128}$ for the whole calculation in the paper.
(For an appropriate initial value of the modification factor,
see the discussions in the references\cite{WangL,TroyerWA}.)

We set the cut off $L$ to be 500.
This limits the accessible temperature range to
$T \gtrsim 0.05 J$.
In the first stage, 
it takes approximately
$6 \times 10^6$ Monte Carlo steps (MCS) before 
the last run with $- \ln \lambda_i < 10^{-5}$ terminates.
In the second stage of simulation,
we fix $f(n)$ and let the random walker travel 
for $1 \times 10^8$ MCS.
The final estimate of the DOS is shown in \figref{fig1}.
\deffig{fig1}{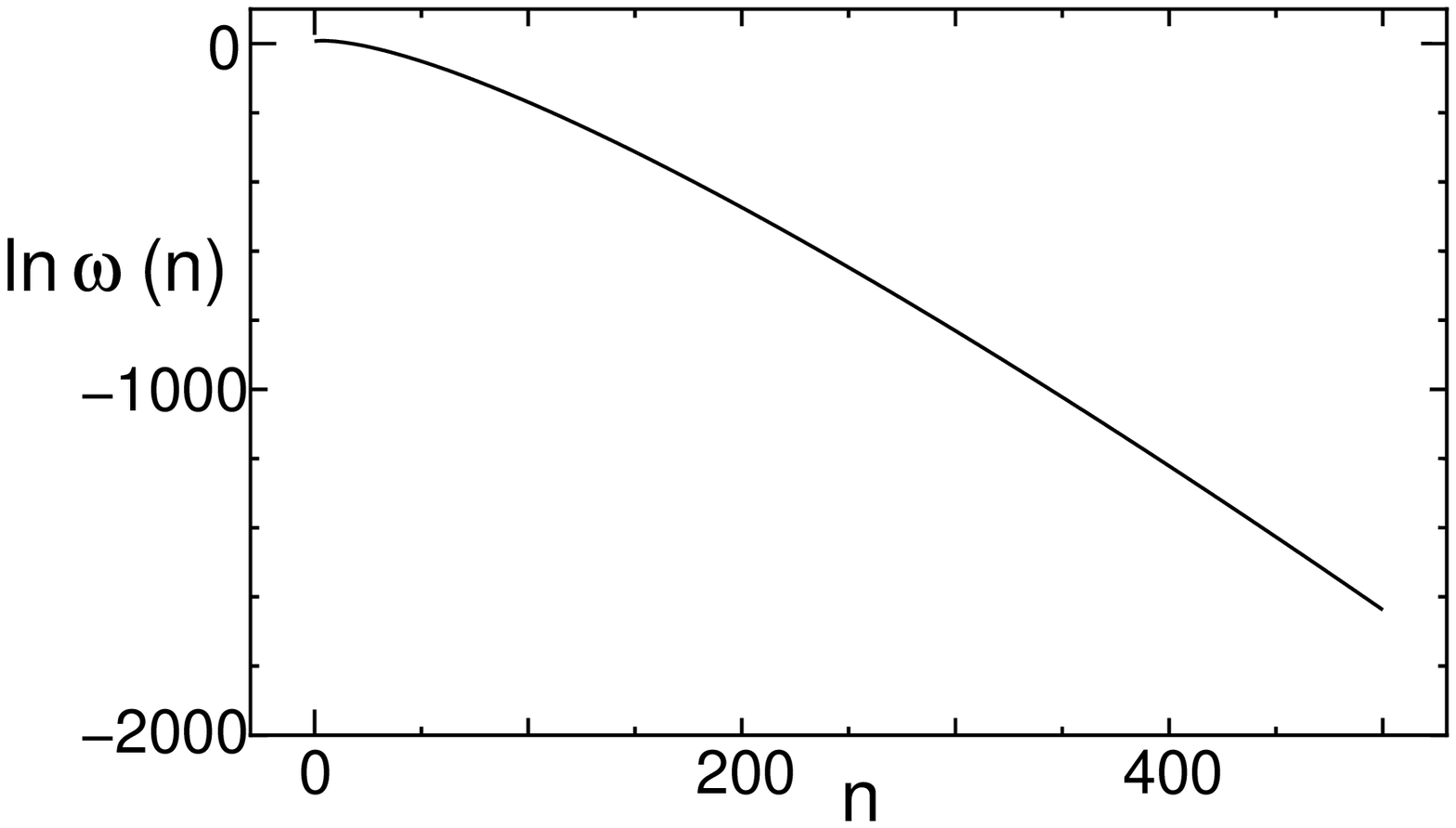}{0.45}
{
The density of states for the series orders for
the antiferromagnetic spin-1/2 Heisenberg chain on ten-site.
The cut off of series orders is $L = 500$.
The first stage of simulation was performed for $6 \times 10^6$
 Monte Carlo sweeps and the second stage was
 calculated for $1 \times 10^8$ Monte Carlo sweeps.
}
By using the DOS in \figref{fig1}, 
canonical (fixed-temperature) averages of the energy, 
the specific heat, the free energy, and the entropy 
are calculated as functions of temperature.

In \figref{fig2}, we show the relative error in the free energy,
$$
\epsilon_F \equiv \left| \frac{F^{({\rm Exact})} (T)
 - F^{({\rm Simulation})}(T)}{F^{({\rm Exact})} (T)} \right| \, .
$$
\deffig{fig2}{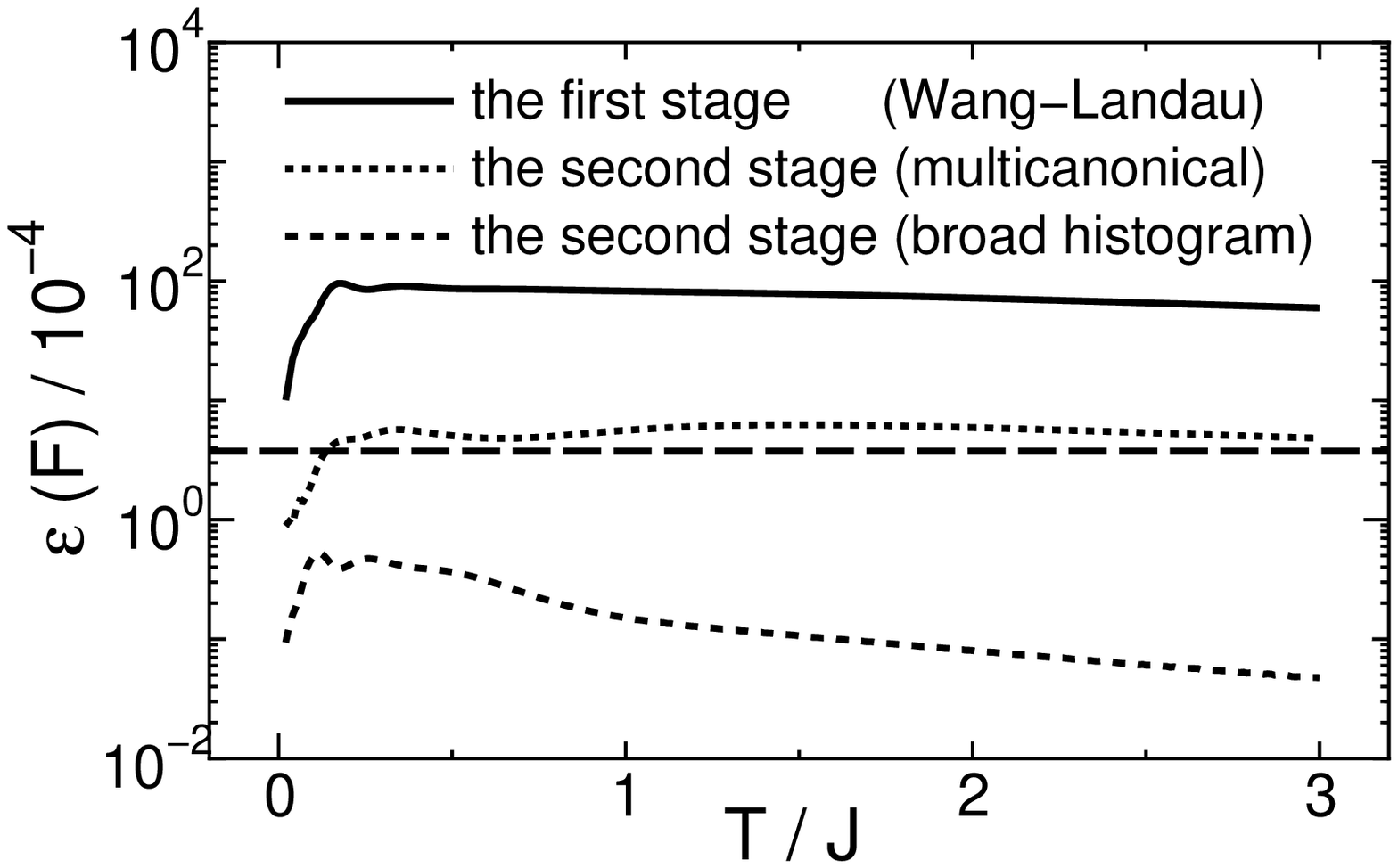}{0.45}
{
Relative errors of the free energy for the antiferromagnetic spin-1/2
 Heisenberg chain on ten-site.
The cut off of series orders is $L = 500$.
The solid line is the result of the Wang-Landau algorithm of the first stage,
which took $6 \times 10^6$ MCS.
The dotted line represents the result of the multicanonical method
 in the second stage of simulation.
The short dashed line is the result by using the broad histogram relation
 in the second stage of simulation.
The dashed line is the result of Troyer {\it et. al.} \cite{TroyerWA}
 that is drawn at the worst value.
}
The top solid line is the one that is obtained based on the 
working estimate of $\omega(n)$ resulting from the first stage
that takes about $6 \times 10^6$ sweeps.
The bottom dashed line represent the one based on the
final estimate of $\omega(n)$ obtained from the second stage
using the estimator \Eq{eq:TheEstimator}.
The middle dotted line also represent the free energy
based on the second stage data but without using the
estimator \Eq{eq:TheEstimator}.
This corresponds to the standard procedure of the
multicanonical simulation \cite{BergN,Lee}.
Namely, the middle curve is based on the naive estimate of the 
DOS using the histogram itself, i.e., the estimate using the
relation $\omega(n) \propto H(n) \omega^{\rm trial}(n)$ 
where $\omega^{\rm trial} \propto (f(n))^{-1}$ is
the trial DOS estimated in the first stage
and $H(n)$ is the number of visits to the value $n$ 
counted during the second stage simulation.
The difference between the middle curve and the bottom one 
is as much as 1 to 2 orders of magnitude, and it clearly shows 
the utility of the estimator \Eq{eq:TheEstimator}.
The horizontal dashed line indicates the level of the
largest error of the previous simulation by
Troyer, Wessel and Alet\cite{TroyerWA} for the
same system with the same size.
The number of Monte Carlo sweeps performed in their
simulation was the same as the second stage of our simulation.
It is natural that their result is about the same position as
the worst (i.e., the highest) point in the middle curve,
since the last few runs in the WL method, where the modification
factor is very close to one, are almost equivalent to the
ordinary multicanonical simulation with fixed weight.

In order to see
the system-size dependence of the efficiency,
we calculate the standard deviation $\sigma$ of the specific heat
per site at temperature $T = 0.5 J$ for various system sizes.
Again, we compare three results: 
(1) the one based on a WL method only with the termination condition
$- \ln \lambda_i < 10^{-8}$ (denoted as ``Wang-Landau'' in the figure),
(2) the one based on the two stage simulation with
the termination condition 
$- \ln \lambda_i < 10^{-5}$ for the first stage and with the naive
estimation of the DOS using the histogram itself
 in the second stage (denoted as ``multicanonical''), and
(3) the same as the above but with the estimator \Eq{eq:TheEstimator}
(denoted as ``broad histogram'').
In (2) and (3), the total number of Monte Carlo sweeps are chosen
to be the same as that is spent in (1) in order to make the
comparison fair.
We choose $L$ and $M$ so that $L/M=50$ is fixed.
With this choice the accessible temperature range is $T \gtrsim 0.05 J$. 
40 independent sets are performed for $M = 8, 12, 16, 20$ whereas
12 for $M = 24$.
The results are shown in \figref{fig3}.
\deffig{fig3}{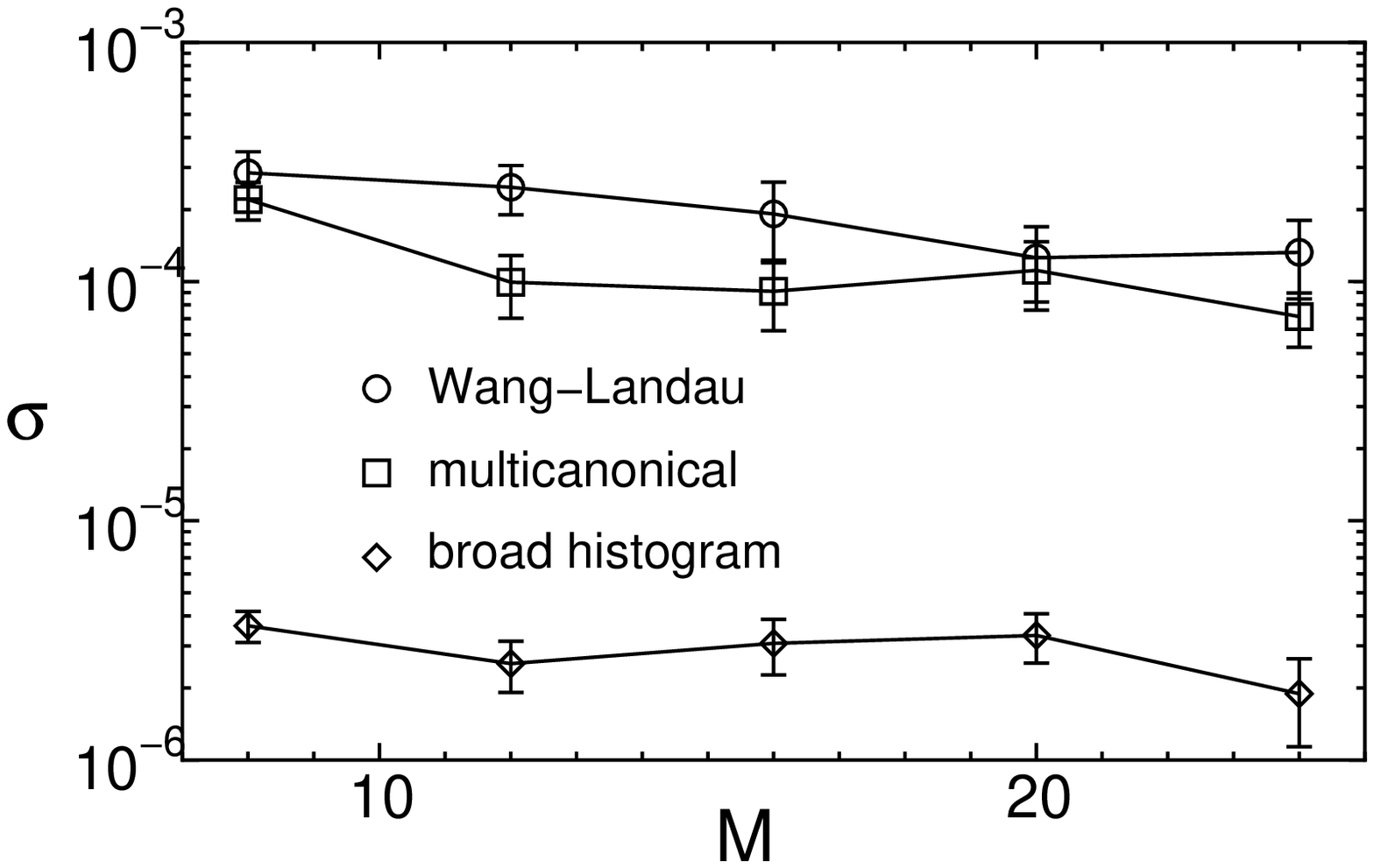}{0.45}
{
The system-size dependences of the standard deviation $\sigma$
 for the specific heat per one site at $T = 0.5 J$
 for the spin-1/2 antiferromagnetic Heisenberg chain.
Three methods were compared:
  the generalized WL algorithm (circles),
 the combination of  the multicanonical method and
 the generalized WL algorithm (squares), and the proposed method (diamonds).
The methods spent the same MCS for each $M$.
We took the $M$ and the $L$ as
 $(M, L) = (8, 400), (12, 600), (16, 800), (20, 1000)$.
The accessible temperature range is $T \gtrsim 0.05 J$.
40 sets of simulation are performed for $M = 8, 12, 16, 20$ and
12 sets for $M = 24$.
}
The system size dependence is small.
The results of the WL algorithm and 
the naive estimate of the DOS based on the two-stage simulation
are close to each other again.
The proposed method outperforms the other method
by more than an order of magnitude in all the cases studied here.


\section{Summary}
Generalizing the histogram methods\cite{JankeK,YamaguchiK,YamaguchiKO}
based on the graph representation,
we have presented a Monte Carlo method for quantum lattice models
that consists of two stages.
For the method's performance, the generalized broad histogram
relation \Eq{eq:BHRelation}, proved for quantum systems, is crucial.
In the first stage, the WL method is employed as in 
\cite{TroyerWA} for obtaining the working estimate of the DOS 
with a short CPU time.
In the second stage, microcanonical averages are computed
and the final estimate of the DOS is obtained from them
through the use of the generalized BH relation.
In both the stages, the update of the states are performed
by loop algorithm.
We have demonstrated that the proposed method yields
very accurate estimates of the DOS in the case of
the $S=1/2$ antiferromagnetic Heisenberg model, which we 
believe to be a typical case that represents many other
cases.
In Ref.~\cite{TroyerWA},
an alternative usage of the extended
ensemble method was proposed in which one
takes the series expansion not in $\beta$
but in other coupling constants.
This type of extension is also available in the present method.

\section{Acknowledgment}
We thank M.~Troyer, S.~Wessel and F.~Alet for comments and 
the exact diagonalization results that were useful for
the error estimation.
We are also grateful to K.~Harada, H.~Otsuka, Y.~Tomita, 
and T.~Surungan for useful comments.
C.Y. thanks T.~Horiguchi, Y.~Fukui, K.~Tanaka,
 K.~Katayama, N.~Yoshiike, and T.~Omori for useful comments.
N.K.'s work was supported by the grant-in-aid (Program No.14540361)
from Monka-sho, Japan.

\end{document}